\documentclass[graybox, envcountchap]{svmult}

\usepackage{mathptmx}       
\usepackage{helvet}         
\usepackage{courier}        
\usepackage{natbib}
\usepackage{makeidx}         
\usepackage{graphicx}        
\usepackage{multicol}        
\usepackage[bottom]{footmisc}

\usepackage{color}        
\input{journalnames.sty}      
\usepackage{hyperref}

\hypersetup{
pdftitle={Kinematically Detected Halo Streams},
pdfauthor={Martin C. Smith},
colorlinks=true,
linkcolor=blue,
citecolor=blue,
urlcolor=blue
}

\usepackage{cite}

\begin{document}

\bibliographystyle{apj}

\title*{Kinematically Detected Halo Streams}
\subtitle{{\small To appear in ``Tidal Streams in the Local Group and
  Beyond'', Astrophysics and Space Science Library, Volume
  420. Springer International Publishing Switzerland, 2016}}

\author{Martin C. Smith}
\institute{Key Laboratory for Research in Galaxies and Cosmology,
  Shanghai Astronomical Observatory, Chinese Academy of Sciences, 80
  Nandan Road, Shanghai 200030, China; email:msmith@shao.ac.cn
}

\maketitle

\abstract{Clues to the origins and evolution of our Galaxy can be found in the
kinematics of stars around us. Remnants of accreted satellite galaxies produce
over-densities in velocity-space, which can remain coherent for much longer
than spatial over-densities. This chapter reviews a number of studies that have
hunted for these accretion relics, both in the nearby solar-neighbourhood and
the more-distant stellar halo. Many observational surveys have driven this
field forwards, from early work with the Hipparcos mission, to contemporary
surveys like RAVE \& SDSS. This active field continues to flourish, providing
many new discoveries, and will be revolutionised as the Gaia mission delivers
precise proper motions for a billion stars in our Galaxy.}

\section{Introduction to kinematic streams}
\label{sec:intro}

At first glance the kinematics of disk stars in the
solar neighbourhood might appear to be a smooth distribution, but upon closer
inspection one can uncover a wealth of structure. The fact that this velocity
distribution is clumpy has been known for over a century. The German astronomer
J.H. von M{\"a}dler, while carrying out observations to measure the
Sun's motion, noticed clumping in the distribution of proper motions
\citep{Madler1846}. This collection of stars moving with the same velocity,
what we now refer to as a ``moving group'', consists of members of the
Pleiades open cluster, including stars several degrees from the centre
of the cluster. This work was build on by \citet{Proctor1869}, who found a
further two moving groups - Hyades and Sirius.

\begin{figure}
\centering
\includegraphics[width=\hsize]{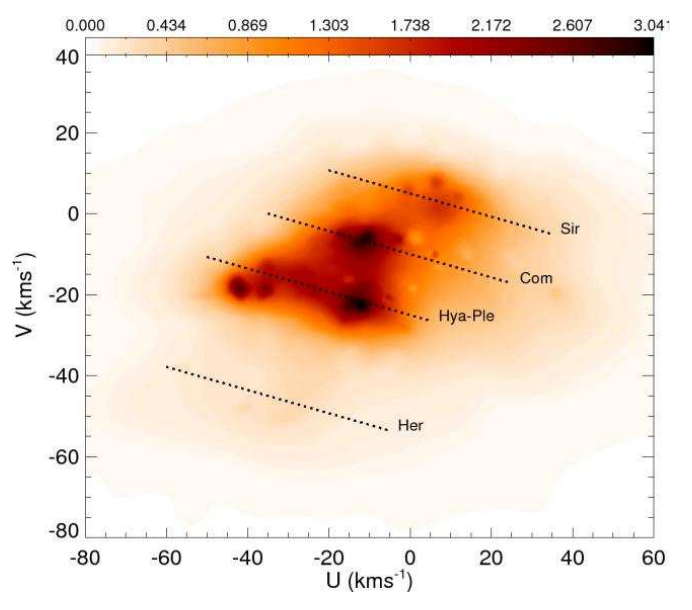}
\caption{The local stellar velocity distribution from a compendium of
observational datasets. The two velocity components correspond to the Cartesian
in-plane velocities, with U increasing in the direction of the Galactic centre
and V in the direction of the Sun's rotation. The approximate locations of four
moving groups are denoted by the dotted lines. Figure taken from \citet{Antoja2010}.}
\label{fig:UV}
\end{figure}

The dissection of the local stellar velocity distribution has told us a great
deal about star formation and Galactic structure. Figure \ref{fig:UV} presents a
contemporary analysis of the velocity distribution, showing that it is rich in
substructure. These moving groups can be attributed to young open clusters
which have not yet dispersed, or could be due to dynamical effects such as
stars trapped at resonance with the Galactic bar or spiral arms \citep[e.g.][]{Dehnen1998,Bovy2010,Sellwood2010,McMillan2011}. However, neither of
these will be explored in this chapter; interested readers are suggested to see
\citet{Antoja2010} for a detailed and comprehensive review on the subject of
moving groups. Instead we will here focus primarily on a third mechanism for
creating moving groups, namely the accretion of extra-galactic
systems.

As can be seen from the other chapters in this volume, we know of many tidal
streams in the stellar halo of the Milky Way. However, more diffuse streams,
which are created if the progenitor is less massive, is on a highly eccentric
orbit, or was tidally disrupted in the distant past, are harder to identify as
spatial over-densities on the sky. They can, however, be discovered by
exploring higher-dimensional space; a stream can that has dispersed in
configuration space, may still remain coherent in phase space. In this
chapter the many successful efforts to find fainter and more ancient accretion
events will be reviewed, starting with the solar neighbourhood (Section
\ref{sec:local}) and moving out to structures in the more distant halo
(Section \ref{sec:distant}).

\section{Local kinematic streams}
\label{sec:local}

\subsection{The Helmi stream}
\label{sec:helmi}

In the 150 years since M{\"a}dler made his initial discovery, many other studies
have identified kinematic substructures. Some of these, such as the Arcturus or
Kapteyn group, have been proposed to be extra-galactic in origin (we will
return to these later). However, the first indisputable accretion remnant was
found by Helmi et al. in their seminal 1999 paper \citep{Helmi1999}, which we
will now discuss.

The first step to identifying clumping in velocity
space is, by definition, obtaining a sample of accurate velocities. Large
surveys have been crucial in propelling many fields in astronomy, from
Kapteyn's visionary international survey of the early 20th century
\citep{Kapteyn1906,Kinman2000,vdKruit2015}, to the Sloan Digital Sky
Survey in the early 21st \citep{York2000}; for a review of many
significant results from such surveys in the field of Milky Way
science, see \citet{Ivezic2012}. One of the most important surveys in
the past 20 years in the field of stellar kinematics was the Hipparcos
mission. This was a European-led satellite mission, launched in 1989
and operated until 1993, which measured the position of stars to
unprecedented accuracy. This enabled a catalogue of proper motions and
parallaxes to be constructed for over 100,000 stars \citep{ESA1997},
although the complex nature of this task meant that the final
definitive catalogue was only published in 2007
\citep{vanLeeuwen2007a,vanLeeuwen2007b}.

Possessing such accurate proper motion data for large numbers of stars enables
statistically meaningful analyses to be done of rare types of stars. For
arguably the first time researchers were able to construct significant samples
of stars from the halo, which remember constitutes only around 0.5 per cent of
the stellar mass in the solar neighbourhood \citep{Juric2008}. With this
sample, one could now begin to search for kinematic over-densities.

The groundwork for Helmi's discovery was laid in the 1999 paper entitled
``Building up the stellar halo of the Galaxy'' \citep{Helmi1999b}. This
paper used a combination of numerical simulations and analytic analysis to
investigate the disruption of satellite galaxies as they are accreted into the
Milky Way, paying particular attention to the consequences for the local
stellar halo. By analysing the disruption of the satellite using action-angle
variables, they studied how the tidal stream evolved over 10 Gyr, and showed
that the system became phase mixed; at a particular location (for example in
the solar neighbourhood) one might detect multiple kinematic over-densities from
the same progenitor. This is because one (large) progenitor can produce a tidal
stream that wraps multiple times around the Milky Way, and can produce numerous
separate clumps in phase space. A simple calculation shows that if the local
stellar halo is entirely made up of debris from 10-100 accreted satellites (of
luminosity 1e7 or 1e8 $\rm L_\odot$) then one would expect to find a few hundred
separate kinematic streams in the solar neighbourhood, although the individual
stars in these streams would not necessarily be clumped in density.

Given the high quality data that was becoming
available at that time, namely the aforementioned Hipparcos catalogue, \citet{Helmi1999} were able to construct a catalogue of 97 metal-poor red giants and
RR Lyrae stars within 1 kpc of the Sun (using data compiled by
\citet{Beers1995} and \citet{Chiba1998} With distances estimated from
photometry (to $\sim 20$ per cent), radial velocities from spectroscopy (to
$\sim 10 ~\rm km/s$) and proper motions mainly from Hipparcos (to a few
mas/yr), then were then able to analyse the phase-space distribution
of this halo sample.

One then needs to determine the optimal space (combinations of
$\rm x$, $\rm y$, $\rm z$, $\rm v_x$, $\rm v_y$, and $\rm v_z$) in
which to identify accretion debris. To do this one normally identifies
integrals of motion, i.e. quantities that are conserved (or
approximately conserved) along an orbit
\citep[see section 3.1.1 et seq of][]{Binney2008}; 
stream stars are thus clumped in these quantities. For example, in a spherical
potential we know that the total energy and the three components of angular
momentum are all conserved along an orbit, leading to four integrals of motion.
However, the Milky Way is not a spherical system, particular in the inner
regions where the contribution of the disk to the total potential is important.
Fortunately it is close to being axi-symmetric (since we do not need to worry
about the Galactic bar, which is restricted to the inner-most few kpc). For an
axi-symmetric system, there are two integrals of motion: the energy, and the
component of the angular momentum parallel to symmetry axis ($\rm
L_z$). It can also be shown that the total angular momentum, although
not precisely conserved, varies only slightly for modest amounts of
flattening and, importantly, shows no long-term evolution.

There are small complications to this picture. First, the stars of a satellite
galaxy do not all follow the same precise orbit, due to the internal velocity
dispersion of the system. This means that stars will have a finite volume in
this space defined by the integrals of motion. Second, the potential of the
Milky Way will undoubtedly be varying over the time-scales under discussion,
for example from the build-up of the disk as gas settles into the equatorial
plane of the Galaxy. This leads to an oft-used term in this field - adiabatic
invariants \citep[see section 3.6 of][]{Binney2008}. These are
quantities that are constant for slowly-evolving potentials,
i.e. potentials that vary on time-scales longer than typical orbital
periods.

\begin{figure}
\includegraphics[width=0.5\hsize,trim=0 100 0 150,clip]{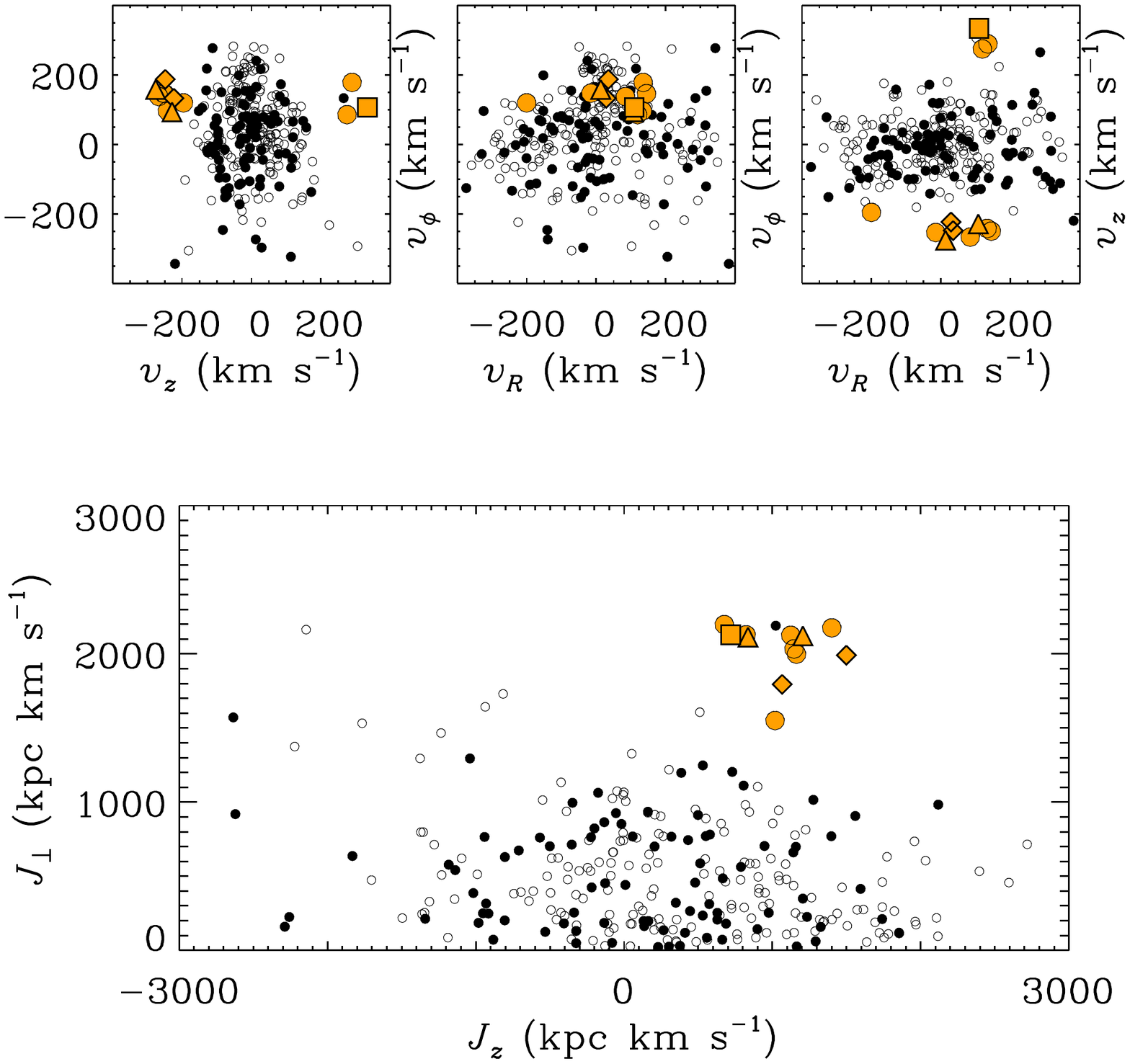}
\includegraphics[width=0.5\hsize,trim=0 0 0 10,clip]{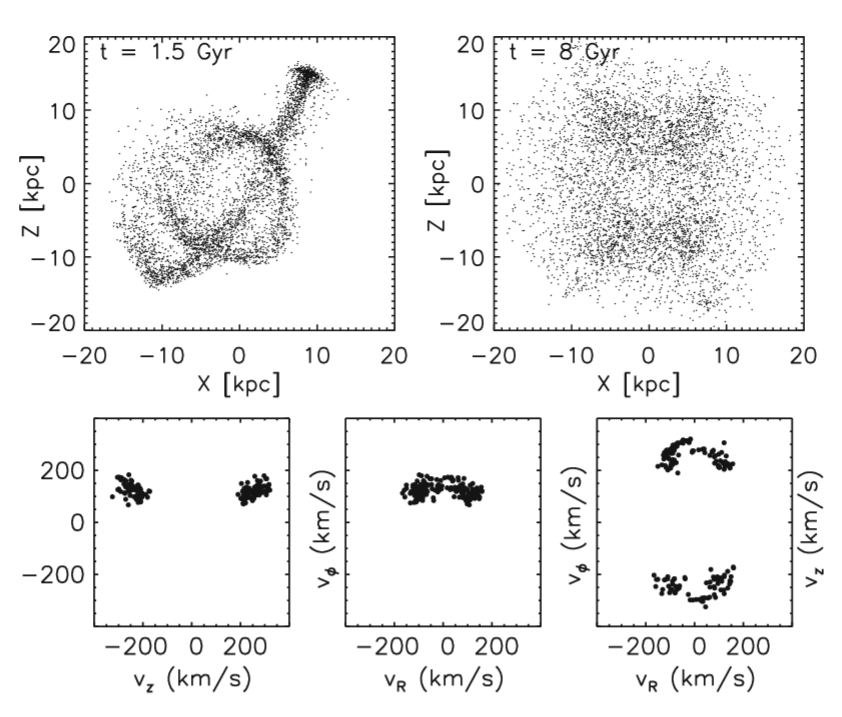}
\caption{The discovery of
kinematic substructure in the solar neighbourhood, from \citet{Helmi1999}. The
left panel shows the observational discovery, which was based on an analysis of
stars from the Hipparcos mission. The lower panel shows the distribution in
angular momentum space (in their notation $\rm J = L$), while the
upper panel shows the kinematics. The right panel shows an N-body
model of the stream, which is almost entirely phase mixed after 8
Gyr. This panel was taken from \citet{Helmi2008}.}
\label{fig:H99}
\end{figure}

For their analysis, \citet{Helmi1999} chose to use the 2-dimensional space of
$\rm L_z$ and $\rm L_\perp = \sqrt{L_x^2+L_y^2}$. They did this as the former is
conserved for an axi-symmetric potential and, since
$\rm L^2 = L_\perp^2 + L_z^2$ and both $\rm L$ and $\rm L_z$ are
approximately constant for stars in a particular tidal stream, the
latter is not expected to vary significantly. These two quantities
were selected because they are trivial to determine for their sample,
unlike the energy which, although conserved,
requires one to assume a model for gravitational potential. The distribution of
$\rm L_z$ and $\rm L_\perp$ is shown in Fig \ref{fig:H99}. For a local
sample of stars $\rm L_z \approx R_\odot . v_\phi$, where $\rm v_\phi$ is
the azimuthal component of the velocity. Since the stellar halo shows no
significant signs of rotation \citep[e.g.][]{Smith2009a}, one would expect that
the distribution should be symmetric about $\rm L_z = 0$; clearly this is not the
case at high $\rm L_\perp$, where there is a significant cluster of stars in the
region ($\rm L_z,~L_\perp$) = (1000, 2000) $\rm kpc.km/s$. The
probability of such a clump of 7 stars occurring by chance is
estimated to be less than 1 per cent. Although their original halo
sample contained only 97 stars, they augmented it with more distant
and more metal-rich stars, obtaining a total of 12 stars that belong
to this overdensity.

They also investigated the velocity distribution of this stars from this system
(top panels of Fig. \ref{fig:H99}; see also Figure 1 of
Chapter 1 which shows a subset of the sample in a slightly different coordinate
system). Note that in Helmi's coordinate system, positive $\rm L_z$
implies $\rm v_\phi$ is in the same direction as Galactic rotation, so
these stars have azimuthal velocities similar to the Sun, but with
very high vertical velocities (as can be seen from the top panels of
Fig. \ref{fig:H99}, $\rm v_z$ is 
around 200 km/s). It is also interesting to see that $\rm v_z$ is split into two
separate clumps -- one moving downward and one upward. This is most-likely a
manifestation of the aforementioned phase-mixing, where one progenitor can
produce multiple streams in phase-space. So although there are two streams,
these stars are all from the same accretion event. Helmi and collaborators
constructed an N-body simulation of this system, which provides a very good
representation of the observed velocity distribution (Fig.
\ref{fig:H99}). This shows how the progenitor can be
almost completely disrupted in configuration space, yet remain coherent in
velocity space. The two separate clumps in $\rm v_z$ are reproduced, as is the
spread in $\rm v_R$. The simulation also allows them to estimate the properties of
the progenitor, which they concluded was likely to be similar in size to the
Fornax dwarf spheroidal.

Subsequent studies have confirmed the Helmi stream and
increased the number of member stars, though somewhat reducing the fraction of
the halo it comprises (for example
\citet{Chiba2000,ReFiorentin2005,Dettbarn2007,Kepley2007}
Later surveys containing fainter stars have
enabled us to identify halo subdwarfs in this system
\citep{Klement2009,Smith2009a}, bringing the total number of members
to over 30 out to distances of a few kpc. The reported fraction of
local stellar halo made of Helmi stream debris varies from reference
to reference, probably due to differences in search techniques, but is
most likely around 5 per cent. The chemistry of 12 of these stars has
been investigated by \citet{Roederer2010}. They find a spread in
$\rm [Fe/H]$ from $-3.4$ to $-1.5$, with detailed abundances similar to that of
the general halo population, concluding that star formation in the
progenitor was truncated before the products of Type Ia supernovae or
AGB stars enriched the inter-stellar medium.

The Kepley reference also introduces a novel way to estimate the age of
accretion. They utilise the fact that once the progenitor has become completely
phase-mixed, the total number of stars in the positive and negative
$\rm v_z$ clumps would be the same. The observed fraction with
positive $\rm v_z$ is actually 28 per cent, which means it is not
entirely mixed. By analysing an N-body simulation
of the accretion (similar to the right-panel of Fig. \ref{fig:H99}), they
can tentatively say that it was likely to have been accreted between 6 and 9
Gyr ago, i.e. not too recently (otherwise the ratio would be more lopsided) and
not too long ago (otherwise the ratio would be closer to unity).

\subsection{Other early discoveries}
\label{sec:early}

As pointed out in the introduction, moving groups have been studied for many
years, but definitively determining their origins is not easy. As a
consequence, finding halo streams in the solar neighbourhood can be a precarious
venture. Over-densities, as their name suggests, can only be identified as an
excess on top of a background population. However, in this case the background
population in question (the Milky Way disk) is far from a simple homogeneous
population; lumps and gradients can often masquerade as coherent structures and
so great care has to be taken in their classification. This problem has
afflicted many potential stream discoveries and, as can be seen in this
chapter, ambiguities still remain for a number of these.

Various streams, such as Hyades-Pleiades, Hercules and Arcturus, are believed to
be Milky Way stars brought together by resonances in the disk (as mentioned in
the introduction to this chapter). Originally identified by Eggen (see
his \citeyear{Eggen1996} review), the broad metallicity or age spread
of these three streams argues against disruption of a disk star cluster as
the origin, and abundance patterns similar to disk stars disfavours the
extragalactic scenario, hence the conclusion of a dynamical origin
\citep[e.g.][]{Famaey2008,Bovy2010,Fuchs2011}. Some of these are more
controversial than others; the Arcturus stream has generated much
debate \citep[e.g.][]{Navarro2004}, but more-recent analyses 
have shown that the narrow velocity dispersion \citep{Bovy2009} and chemical
inhomogeneity \citep{Williams2009,Bensby2014} favor a dynamical
origin. Making definitive claims about the origins of such systems is not easy.

Another stream that has been studied extensively is the Kapteyn moving group,
again named by Eggen. The name comes from the fact that the velocities of this
group are similar to that of Kapteyn's star, which is a well-known very high
proper motion star only 4 pc away from the Sun. This group is on a
mildly retro-grade orbit, moving at around $-50$ km/s.

\begin{figure}
\centering
\includegraphics[width=0.6\hsize]{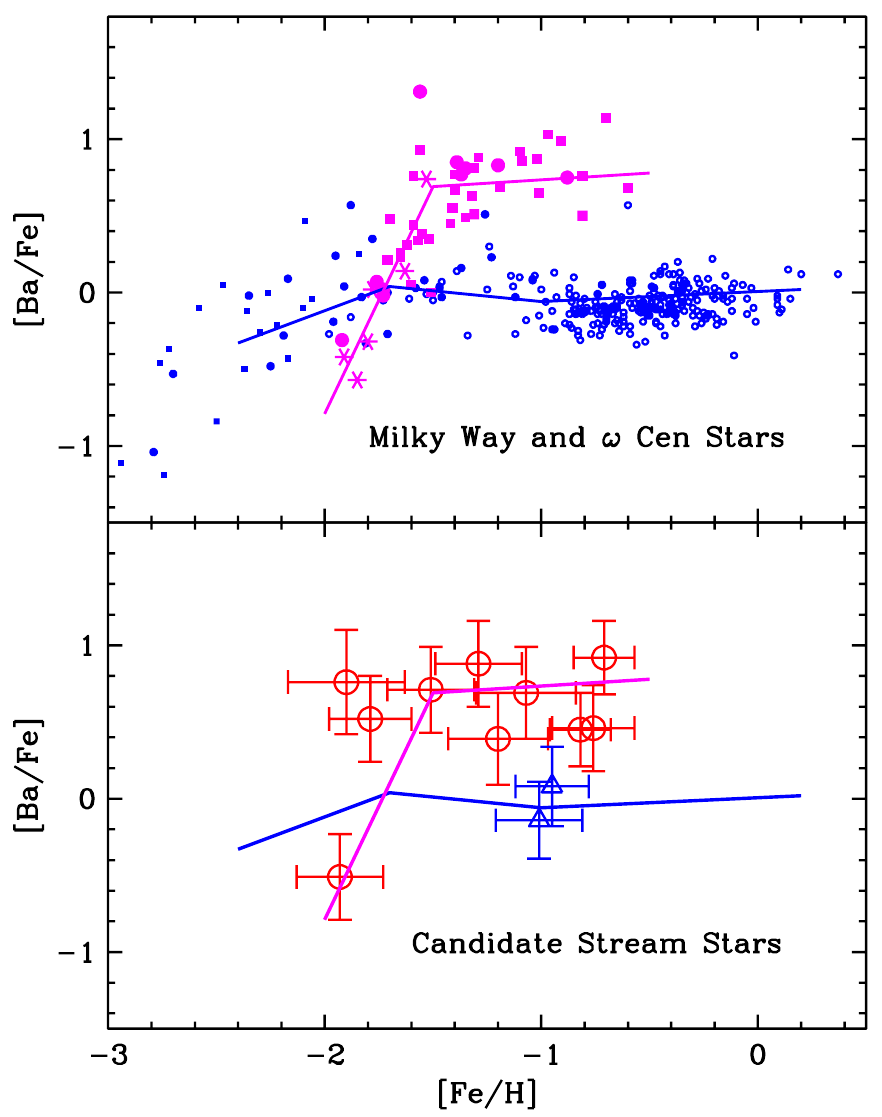}
\caption{Top panel: The distribution of $\rm [Ba/Fe]$ vs
$\rm [Fe/H]$ for Milky Way stars (blue points) and a characteristic locus (blue
line), overlaid with the same for $\omega $Cen stars (magenta points and line),
where data for both are taken from the literature. Bottom panel: the
distribution of barium abundances for 10 stars in the retrograde kinematic
stream discussed in Majewski et al. (\citeyear{Majewski2012}; red),
plus two retrograde stars not in the stream (blue). This retrograde
stream is similar to the Kapteyn stream and is potentially associated
to Omega Cen, a hypothesis which is supported by the close similarity
in chemistry between the cluster stars in the top panel and the stream
stars in the bottom. Taken from \citet{Majewski2012}.}
\label{fig:omegacen}
\end{figure}

Quite early it was speculated that this retro-grade
group might be related to the globular cluster Omega Cen
\citep[e.g.][]{Eggen1978},
which is on a similar orbit. This cluster is interesting as it appears to
possess a broad spread in both chemistry and age, indicating that it could be
the core of a disrupted dwarf galaxy \citep{Lee1999}. This might suggest that
this retrograde moving group is in fact the stripped remnants of the galaxy. A
number of works have pursued this hypothesis
(see \citealt{Majewski2012} for a nice overview of the current situation),
including a detailed spectroscopic
analysis of 16 stars by \citet{WDB2010}. This latter reference
confirms the similarity in chemistry between the cluster and the moving group,
but provocatively questions whether the group may in fact be material stripped
from the host galaxy of Omega Cen as it was disrupted. The reason is that one
would need a significant amount of dynamical friction to bring the system onto
its current orbit, something which would not be possible for an object as small
as Omega Cen. However, the authors acknowledge that
their hypothesis is currently
unproven; if they are correct that Omega Cen and the
Kapteyn moving group do not originate in the same dwarf galaxy progenitor, then
why the close similarity in chemistry and why the age spread in the cluster?
Clearly more work needs to be done to address this question and fully
understand the connection, especially given that a large fraction of the inner
halo could belong to this moving group 
(see, for example, \citealt{Majewski2012}; Fig. \ref{fig:omegacen})
and hence it could have played a significant role in the early
development of our galaxy.

\subsection{Streams in the Geneva-Copenhagen survey}
\label{sec:gcs}

\subsubsection{Pieces of the puzzle}
\label{sec:puzzle}

Although Hipparcos was (and still is) clearly a hugely important resource for
studying the kinematics of stars in the solar neighbourhood, it was hampered by
a lack of radial velocities. \ What was needed was a systematic survey of
radial velocities, providing a sample with a well-defined selection function.
The Geneva-Copenhagen survey
\citep{Nordstrom2004,Holmberg2007,Holmberg2009,Casagrande2011} produced
radial velocities and metallicities for
13,000 nearby stars. Combining these with proper motions and (for the majority
of stars) distances from the Hipparcos mission, resulted in a complete sample
of F- and G-type dwarfs within 40 pc and a larger, magnitude-limited sample to
around 200 pc. Although this is a relatively local sample compared to the one
analysed in Helmi's \citeyear{Helmi1999} study, and contains only a
handful of halo stars, the accurate 6D phase-space information allows
a detailed analysis of the orbital properties of the stars.

\begin{figure}
\centering
\includegraphics[width=0.6\hsize]{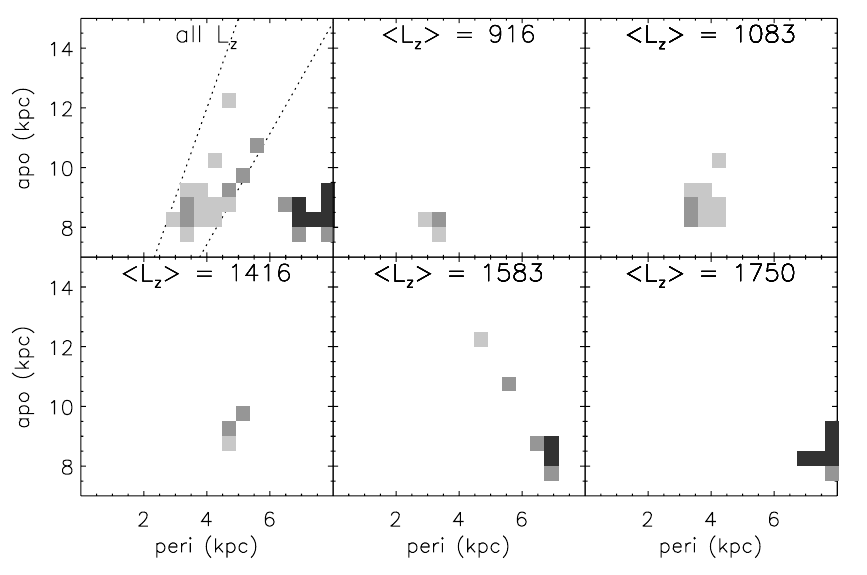}
\caption{Distribution of overdense cells in the
apo-centre/peri-centre/angular-momentum (APL) space. Each panel shows the
distribution of stars in the space of apo- and peri-centre (for different cuts
in the L\_z angular momentum). Darker colours correspond to higher statistical
significance of an excess compared to the smooth background. Notice the group
of stars confined within the dotted lines (corresponding to eccentricities
between 0.3 and 0.5). As L\_z increases, the orbits move outwards in radii.
Taken from \citet{Helmi2006}.}
\label{fig:apl}
\end{figure}

In addition to the Helmi stream, which is the most prominent in
the local stellar halo, a number of other candidate streams have been found.
Helmi continued her analysis using this Geneva-Copenhagen survey
\citep{Helmi2006}, choosing a different space to search for accretion
debris than in their \citeyear{Helmi1999} study. This time they chose
three parameters, L\_z (which, as mentioned above, will be conserved
in an axi-symmetric system) and the peri- and apo-centre distances (the
minimum and maximum distance that the star's orbit comes to the
Galactic centre). We will refer to this as the APL space (Fig. \ref{fig:apl}).
Using simulations they showed that disrupted satellites remain reasonably
localised in this space, with the extent depending on the initial size
of the satellite. When projected into the 2D space of apo- versus
peri-centre distance, debris remain contained within a band of constant
eccentricity, implying that the stars approximately retain the eccentricity of
their progenitor's orbit. Moreover in this space one can see the different
kinematic streams forming as the satellite becomes phase mixed, although the
accuracy required to determine this is not feasible with current datasets.

One clear drawback of this method is that it relies on a knowledge of the
gravitational potential in order to calculate apo- and peri-centre distances.
However, as the sample under investigation is such a small volume, the
potential is approximately constant and hence the orbital parameters are
determined by the kinematics, rather than their location. This means that
although the apo- and peri-centre distances may be slightly off due to an
incorrectly chosen potential, this will result in a systematic shift in this
space and will not work to smear out any coherent features.

By comparing the density of stars in this 3D space to
Monte Carlo realisations of a smooth model, they were able to show that their
distribution is significantly more structured than what one would expect from a
smooth model. They went on to identify a group of
stars with eccentricities around 0.4. This value is clearly inconsistent with
the thin disk, although thick disk stars may posses such high eccentricities.
Folding in metallicities appears to show that this group can be divided into
three separate accretion events, with different trends in $\rm v_z$ and age.

The process of determining exactly how many distinct groups are contained in
such a discovery is complex and open to interpretation. High-resolution
follow-up studies have analysed the elemental abundance patterns of these stars
and suggested that there may actually be only two distinct groups
\citep{Stonkute2012,Stonkute2013,Zenoviene2014}; the common abundance
patterns suggests that Groups 2 \& 3 from the original
\citet{Helmi2006} paper may come from the same progenitor. Tentative
ages for these stars show two populations, a relatively metal-rich one
of around 8 Gyr and a more metal-poor one of around 12 Gyr, but they
acknowledge that more work needs to be done as ages are notoriously
difficult for main-sequence stars. They conclude by speculating that
the similarity between the chemical composition of stars in these two
kinematic groups and in the Milky Way's thick disk suggests that the
progenitor of this system may be related to the formation of the thick
disk.

The original authors revisited these streams in \citet{Helmi2014}, obtaining
high-resolution for even more stars in this eccentricity range (0.3-0.5).
Although they no longer focus on the division proposed in their original paper,
they show that the properties of stars in this eccentricity range are not
homogeneous, with an apparent division around ${\rm [Fe/H] = -0.4}$ dex.

\subsubsection{Arifyanto \& Fuchs}
\label{sec:arifyanto}

In the same year as the Helmi et al. were mining the
Geneva-Copenhagen survey, a group in Germany \citet{Arifyanto2006}
were searching for accretion debris in a sample of 742 stars based on
a catalogue from \citet{Carney1994}. This resulted in the detection
of a number of streams, but none are believed to be due to accretion;
three had been previously identified (Hyades-Pleiades, Hercules and
Arcturus; discussed above) and there was a new discovery, again 
not conclusively extragalactic in origin \citep{Ramya2012}.

However, despite not identifying any bone-fide accretion remnants, this paper
warrants discussion here for its approach to identify nearby streams. Using a
Keplerian approximation for orbits developed by \citet{Dekker1976},
they determine proxies for the total angular momentum and
eccentricity, along with a third parameter corresponding to the
inclination of the orbit (see \citealt{Klement2010} for a 
review of this method). They argue that this three dimensional space is ideal
for dissecting velocity space, as evidenced by their success at finding the
above disk streams. This method is simple to apply and unlike the previous APL
approach does not require orbits to be calculated, but it is not perfect as it
relies on the aforementioned Keplerian approximation that only holds for
spherical potentials. Despite these drawbacks it has proved a popular approach
that has been used by numerous authors, mainly by employing the proxy for the
eccentricity ($\sqrt{U^2 + 2 V^2}$). Although the approximations behind this
proxy break down for high eccentricities, overdensities should remain coherent,
and so the simplicity of this approach has resulted in its widespread use.

Another early paper to adopt this technique was
\citet{Dettbarn2007}, who analysed the sample of \citet{Beers2000} and found a
number of candidate halo streams, including existing groups (such as the Helmi
stream) and some new ones (named, $\rm S_1$, $\rm S_2$, and $\rm S_3$).

\subsection{The modern era}
\label{sec:modern}

\subsubsection{The RAVE survey}
\label{sec:rave}

It should be evident by now that progress in
understanding the formation and evolution of our galaxy rests on large surveys.
We will now discuss two influential surveys in this
field from the past decade: the RAdial Velocity Experiment (RAVE) and the Sloan
Digital Sky Survey (SDSS). Both of these have led to numerous discoveries, as
can be seen from this and many other chapters in this volume.

Taking a kinematic census of large populations of the
Milky Way requires both photometric and spectroscopic surveys. Many proper
motion catalogues have been assembled, often by combining various existing
ground-based astrometric surveys (e.g. the Hipparcos catalogue, UCAC4, PPMXL,
etc, etc). Spectroscopic surveys are in some sense rarer as it requires
significant telescope time to amass spectra for
hundreds of thousands of stars. One of the largest such surveys is the RAVE
survey \citep{Steinmetz2006}, which began in 2003 on the 1.2m
UK Schmidt Telescope at
the Anglo-Australian Observatory. Despite not being particularly new (it was
around 30 years old when the survey started) or having a large aperture, it had
a number of important assets: the field of view was very large (6 degree
diameter field); it had relatively high multiplexing capabilities (150 fibers
per plate); and, crucially, for most of the survey it had dedicated use of the
telescope. The combination of these factors allowed RAVE to efficiently produce
large catalogues of radial velocities and abundances, resulting in a final
catalogue of around 500,000 stars in the Southern hemisphere with magnitude
range 9 {\textless} I {\textless} 12 \citep[the fourth public data
release is presented in][]{Kordopatis2013}. By concentrating
on a narrow window around the calcium triplet region (8410--8795 A) with
reasonable resolution (R \~{} 7500), the radial velocity accuracy is excellent,
at around a couple of km/s \citep[see Fig. 29 of][]{Kordopatis2013}.
Furthermore, stellar parameters can be
derived at a level of around a few tenths of a dex for log(g) and metallicity
\citep[see Table 2 of][]{Kordopatis2013}.
. For the higher signal-to-noise spectra it is
even possible to estimate abundances for a number of individual elements, such
as Mg, Al, Si, Ca, Ti, Fe, and Ni \citep{Boeche2011,Kordopatis2013}.
If we are interested in learning about the kinematics of stars, the final piece
of information is distances, which can be estimated by combining the observed
stellar parameters with stellar models. This was first undertaken by
\citet{Breddels2010} and subsequently refined by \citet{Zwitter2010}
and \citet{Binney2014a}.

With all this information in hand, RAVE has proved to
be a hugely important survey for understanding the kinematics of the
solar-neighbourhood, for example \citet{Binney2014b}. In terms of streams, the
first attempt to investigate this was \citet{Seabroke2008}. By looking at
stars from a couple of surveys, including RAVE, they came to the conclusion
that there were no massive streams (with hundreds of
stars) present in the dataset. They contrasted this
to the situation in the outer halo (see Section \ref{sec:distant} of
this chapter, as well as other chapters in this volume), where the
system is clearly far from relaxed and contains a number of streams
and substructures. The results here are in agreement with previous
studies, for example the work of \citet{Gould2003}, which 
used a sample of 4,000 halo stars with proper motions to conclude that no more
than 5 per cent of the local halo can be made up of a single stream. At first glance
this may appear to conflict with some estimates for the density of the
\citet{Helmi1999} stream discussed above, but one should note that their
discovery is made up of two separate kinematic
streams due to the extensive phase-mixing (Fig. \ref{fig:H99}).

\begin{figure}
\centering
\includegraphics[width=\hsize]{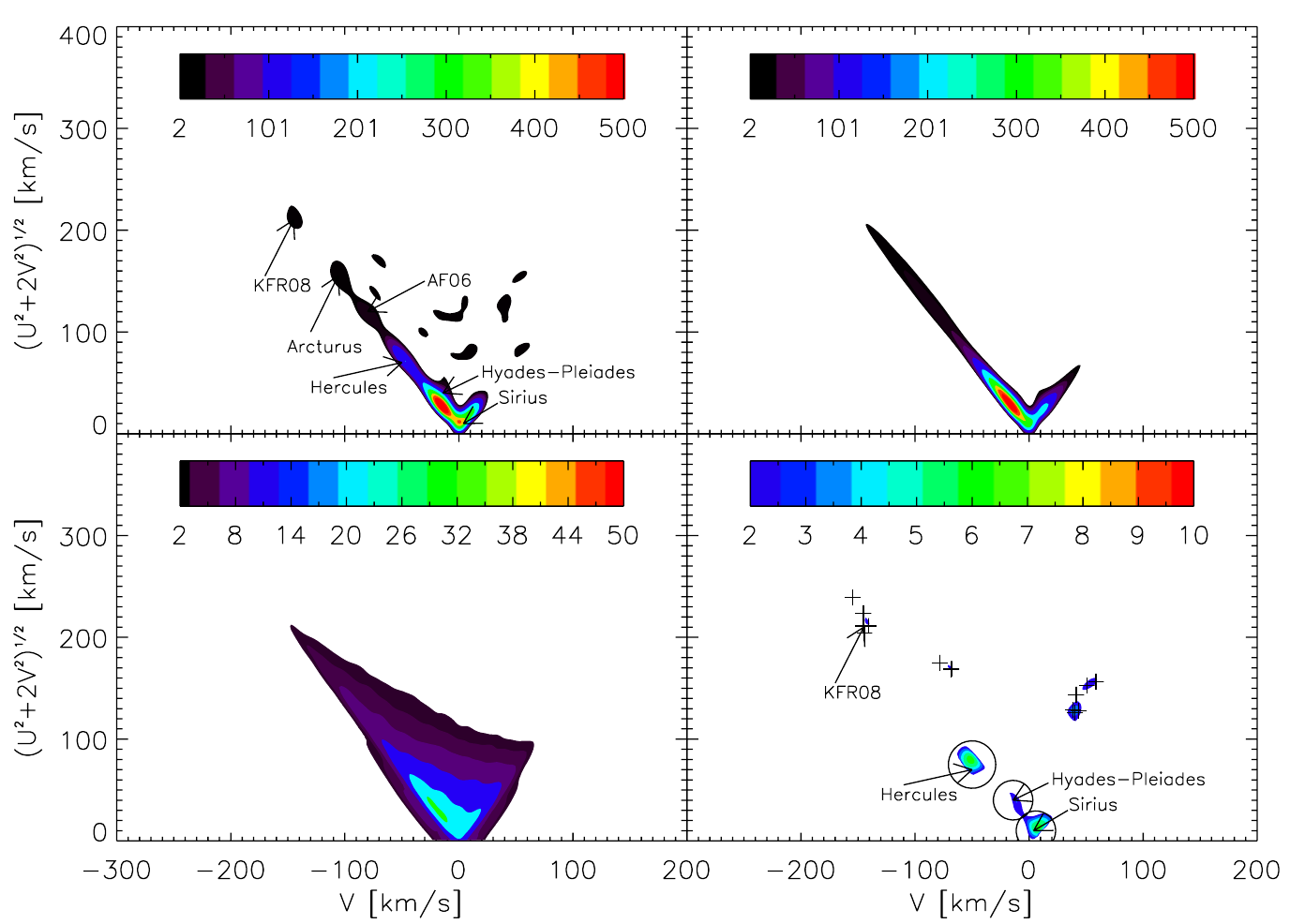}
\caption{Distribution of kinematic over-densities in RAVE, as
presented by \citet{Klement2008}. The upper-left panel shows the
full distribution of the observed dataset, which can be compared to
the expected distribution from a smooth model (upper-right) and its
corresponding dispersion (lower-left). By subtracting the data from
the model, normalised by the dispersion, one obtained the distribution
of overdensities (lower-right). This reveals known substructures, such
as Hercules, and new ones, such as KFR08. This figure is taken from
the later analysis presented in \citet{Klement2011}.}
\label{fig:klement}
\end{figure}

The first attempt to systematically search for
small-scale clumping in velocity space in the RAVE survey was undertaken by
\citet{Klement2008}, who adopted the Keplerian approximation described in the
previous section. One issue here is that 6D phase-space is required and, back
in 2008, distances has not yet been estimated for RAVE stars. To overcome this
issues they made the naive assumption that the sample would be dominated by
main-sequence stars and applied a simple monotonic colour-magnitude relation
suitable for dwarf stars. Due to the bright magnitudes that RAVE probes, this
assumption is rather weak and it was subsequently found that around half of the
stars are likely to be giants or sub-giants \citep[e.g.][]{Binney2014a}. They
later rectified this issue \citep{Klement2011}, basically finding that their
results were unchanged. Despite this problem in their original paper, they were
still able to recover a number of existing streams, calculating their
significance by comparing to expectations from a smooth background
model (Fig. \ref{fig:klement}). One new stream (dubbed ``KFR08'') was also detected at a
level of 3-sigma. Despite 
being relatively metal-rich (${\rm -1 < [Fe/H] < 0}$) and on a
pro-grade orbit (with v\_phi of around 160 km/s), the high vertical velocity
(vertical dispersion being around 100 km/s) indicates that this unlikely to be
associated to the disk and is therefore believed to be the remnant of an
accreted satellite. This was later analysed by \citet{Bobylev2010} in a
sample constructed from Hipparcos stars with accurate trigonometric
parallaxes \citep{vanLeeuwen2007a}
and metallicities and ages \citep{Holmberg2007,Holmberg2009}; by identifying
additional giant stars in this stream they were able to identify the main
sequence turnoff and, through isochrone fitting, determining this is likely to
be a very old stream (likely 13 Gyr in age). We will return to this stream
briefly in the following section, as it is was subsequently confirmed using an
independent data set.

\begin{figure}
\centering
\includegraphics[width=\hsize]{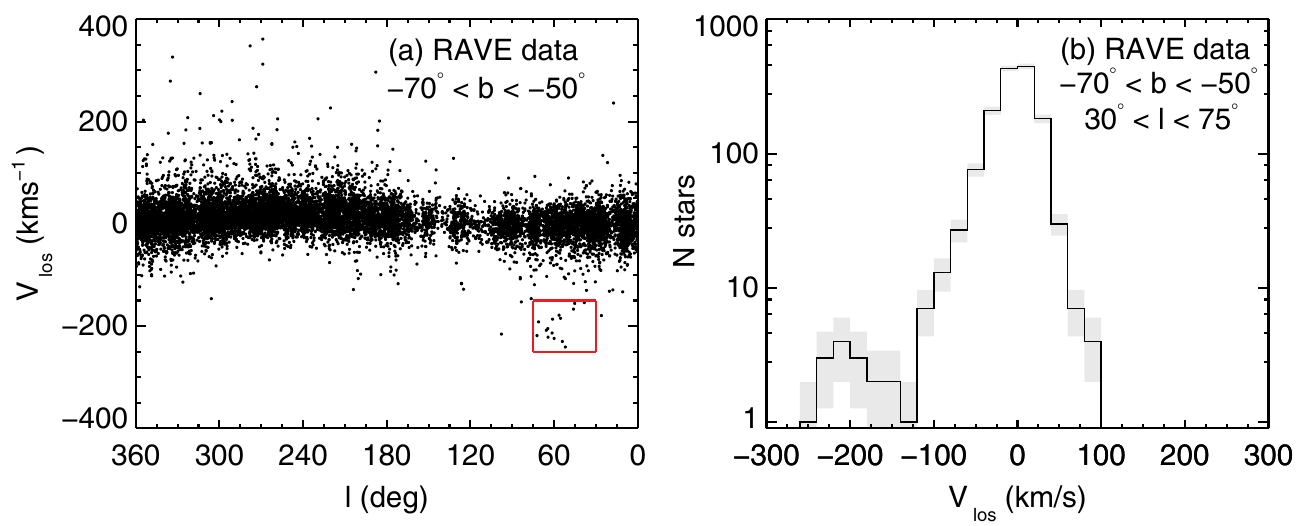}
\caption{These plots, from Williams et al.
(2011), show the disputed Aquarius stream/group. Despite the strong signature
  in kinematics (which we see localised at $\rm v_{los} \sim -200$ $\rm km/s$),
  it is still argued whether this is the remnant of a disrupted
  globular cluster \citep{WDB2012} or just a perturbation in the disk
  \citep{Casey2014}.}
\label{fig:aquarius}
\end{figure}

As discussed above, determining the origin of moving groups can be tricky due to
the complex nature of the background in which they reside (i.e. the Milky Way
disk). This difficulty is highlighted by another candidate stream identified by
the RAVE survey, entitled the Aquarius Stream \citep{Williams2011}. Is was
discovered through its radial velocity offset from the surrounding population
(Fig. \ref{fig:aquarius}) -- a cluster of 15 stars localised on the sky
and with very similar metallicities. This old stream was not connected
to any other existing structures and they concluded that it was likely
recently stripped material from a globular cluster or dwarf
galaxy. However, despite being relatively close by (most of the stars
are within a few kpc), uncertainties in the distances and proper
motions mean that the clumping in angular momentum space is inconclusive.

At this point detailed abundances are required to
definitely determine the origin of this stream. Two teams took up this task,
one led by \citet{WDB2012} using the AAT telescope and one led by \citet{Casey2014} using the Magellan Clay telescope. Unfortunately, instead of
clarifying the issue, these studies raised more questions than they answered.
The first analysis \citep{WDB2012} looked at six member stars,
concluding that the exceptionally tight metallicity spread ($\rm
\sigma([Fe/H]) = 0.1$ dex) and abundance ratios unambiguously show that
the Aquarius stream is a disrupted globular cluster.
The second analysis \citep{Casey2014}, on the other hand, looked at five
stars and unambiguously found
that this system cannot be a disrupted globular cluster, giving their paper the
unequivocal title ``The Aquarius comoving group is not a disrupted classical
globular cluster''.
After finding a wide metallicity spread ($\sigma \rm [Fe/H] = 0.4$
dex) that is inconsistent with a globular cluster and abundance ratios
that are inconsistent with a dwarf galaxy, they conclude that the
stars are indistinguishable from the Milky Way field population.
The coherence
in their kinematics may be a result of the perturbation of the disk when a
satellite fell in, but this is just conjecture and their main conclusion is
that the chemistry of these stars clearly show that they are not the accreted
remnants of either a globular cluster or dwarf galaxy. The fact that any two
studies disagree is not surprising. What IS surprising is that one cannot
easily explain away this discrepancy on the grounds of small number statistics,
since these two analyses have four stars in common! So the fact that very
different conclusions have been reached is hard to reconcile, although the
higher signal-to-noise of the \citeauthor{Casey2014} study certainly
works in their favour and they consequently argue that the
\citeauthor{WDB2012} metallicities may be inaccurate. To conclude,
the nature of the Aquarius stream/group is still open for debate.

\subsubsection{The SDSS survey}
\label{sec:sdss}

In terms of studies of the Milky Way, another major survey in the past
decade has been the Sloan Digital Sky Survey (SDSS). The photometric
part of the survey has led to the discovery of a host of new streams
and dwarf galaxies, including a new population ultra-faint dwarf
galaxies (see \citealt{Belokurov2013} for a recent review). Furthermore it has
spurred on many different analyses on the state of the disk both
through its photometry and spectroscopy (for example, see review
articles by \citealt{Ivezic2012} and \citealt{Smith2012}).

The sub-project dedicated to stellar spectroscopy,
called the Sloan Extension for Galactic Understanding and Exploration (SEGUE;
\citealt{Yanny2009}), was included in SDSS-II and SDSS-III and collected spectra
for nearly 350,000 stars. There were also a large number of stars targeted in
the other sub-projects (over 300,000) and although they have a very
complex selection function, these stars are still
useful for studies of the Milky Way. These SDSS
spectra were taken at a much lower resolution than RAVE (with R
$\approx$ 2000) but with a
much broader wavelength coverage (3800-9200 A). With
this data the SDSS team were able to measure stellar parameters to a precision
similar to that of RAVE, namely log(g) and [Fe/H] to around 0.3 dex \citep{Lee2008}. Alpha elements have also been obtained \citep{Lee2011}, but of course
these are harder to determine and, at the time of writing (i.e. Data Release
12), are yet to be
officially released as part of the SDSS survey. Despite not being officially
released, a number of authors have used these alpha-element abundances, for
example measuring the abundance ``knee'' of the Sagittarius stream \citep{deBoer2014}. The spectroscopy has also led to a variety of papers on the Milky
Way disk, most provocatively the series of papers led by Jo Bovy arguing that
the thick disc is not a separate entity from the thin disk \citep{Bovy2012}.
The SDSS survey is still ongoing, with the stellar spectroscopy now undertaken
by the ambitious APOGEE project, which will be discussed later in Section \ref{sec:future}.

The search for phase-space substructures in SDSS was undertaken by \citet{Klement2009} and \citet{Smith2009a}, using complimentary datasets and
techniques. \citeauthor{Klement2009} took the seventh data release
from SDSS and applied
the same methods as used during their search for overdensities in the RAVE
survey. Since the SDSS stars are much fainter than the RAVE stars, proper
motions are generally less precise. Therefore in order to reduce the
corresponding uncertainties in tangential velocities, they decided to focus on
stars within 2 kpc of the Sun. They were able to confirm their candidate
overdensity from the RAVE survey (the KFR08 group, discussed above), further
breaking this into two separate groups, one of which is likely to have a disk
origin (R1) and one of which is likely accretion debris (R2). They also confirm
one of the previously mentioned streams from \citet{Dettbarn2007}, adding
more members and arguing that this is actually smeared out in phase-space,
resulting in multiple clumps in their analysis. As with most works, the Helmi
stream was found with very strong significance. Finally, two additional
candidate halo streams were identified (C1 \& C3).

At the same time as \citet{Klement2009} were working
on their paper, \citet{Smith2009a} were working on an independent study using
SDSS data. Their approach was complimentary in that it was able to probe much
greater distances thanks to precise proper motions from \citet{Bramich2008}.
These proper motions were constructed using the 250 sq. deg. Stripe 82 region
of SDSS, which was repeatedly monitored primarily for the purpose of detecting
supernovae. However, the multiple epochs are also ideal for constructing proper
motion catalogues. The fact that the proper motions come from a single survey
avoids the problem of cross-matching data from separate
telescopes with often very different conditions and imaging
systems, significantly reducing the systematic errors. Despite the relatively
short baseline (seven years), proper motions were measured to a few mas/yr
precision, even at faint magnitudes. \citet{Koposov2013} updated the
catalogue using improved techniques, reducing the systematic
errors even further and obtaining a precision of around 2 mas/yr. The stunning
precision enabled the authors to detect the proper motion of the Sagittarius
stream, even at a distance of 30 kpc.

\begin{figure}
\centering
\includegraphics[width=0.45\hsize]{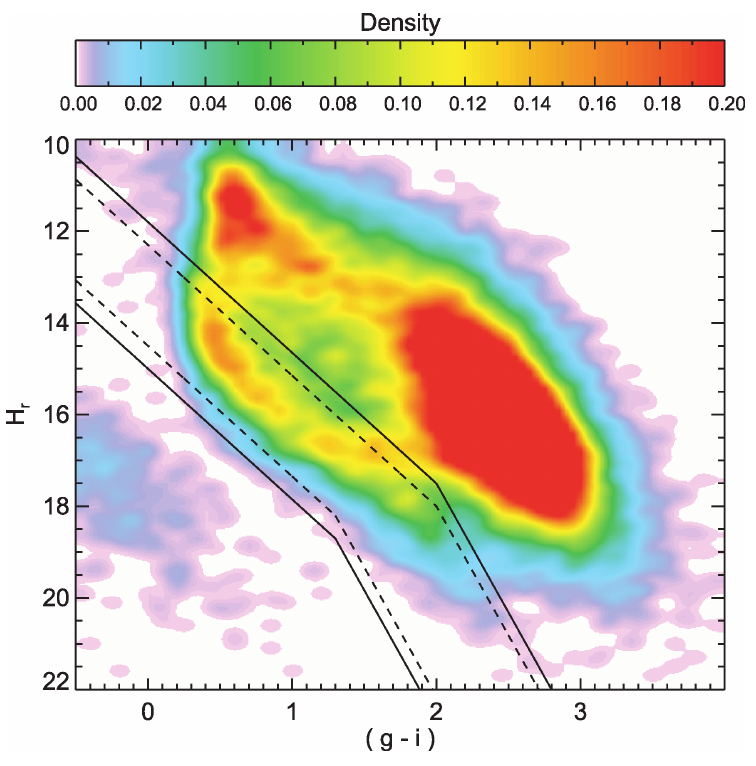}
\includegraphics[width=0.45\hsize]{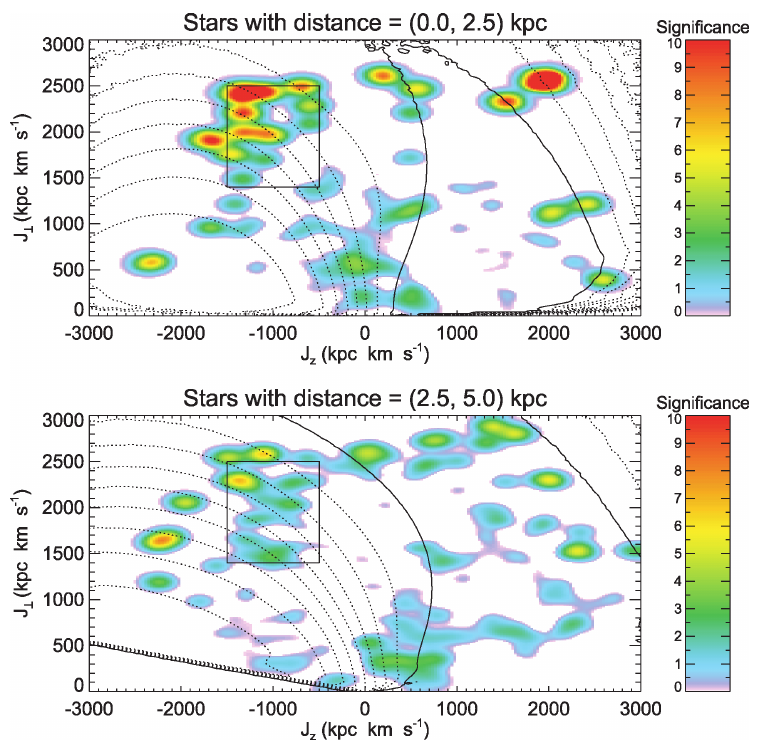}
\caption{Analysis of kinematic overdensities in SDSS subdwarfs. The left panel shows how
the subdwarfs can be selected from a reduced proper motion diagram. The
subdwarfs, which are located within the dotted lines, separate from the disc
sequence owing to their fainter intrinsic magnitudes and greater velocities.
The two right panels show the resulting distribution of subdwarfs in angular
momentum space, after subtraction of a smooth halo model. Note the prominent
Helmi stream, located around $\rm J_z = -1000$ and $\rm J_\perp =
+2000$ kpc $\rm km/s$. The
dotted lines in these panels denote the detection efficiency, which has to be
considered as these stars are kinematically selected. Figure taken from \citet{Smith2009a}.}
\label{fig:smith}
\end{figure}

These accurate proper motions allowed \citeauthor{Smith2009a} to
investigate halo stars out
to 5 kpc, with accuracy around 30-50 km/s for each component of the velocity.
The halo stars were identified using a reduced proper motion diagram
(Fig. \ref{fig:smith}).
This technique, which uses the proper motion as a proxy for distance, is often
used to separate (nearby) dwarfs from (distant) giants, but is a powerful
technique to separate halo stars from disk stars, as halo stars are faster
moving and hence have higher proper motions for a given distance. Although this
selection is kinematically biased, it is easy to correct for this as the cut in
reduced proper motion corresponds to a cut in tangential velocity; from a
simple model one can calculate the detection efficiency for stars of a given
velocity without accurate distances or luminosity functions. An additional
factor which~helps to separate out halo stars is that metal-poor stars are
bluer, meaning that (for a given colour) metal-poor stars are intrinsically
fainter than metal-rich~ones by as much as a couple of magnitudes (see, for
example, equation A2 of \citealt{Ivezic2008}; see also \citealt{Bochanski2013}).
Once the halo stars were identified photometrically from the reduced proper
motion diagram, they were cross-matched with SDSS spectroscopy to provide
metallicities and velocities. Distances were estimated by combing spectroscopic
metallicities with the photometric distance relation of \citet{Ivezic2008},
amended slightly to include a minor correction as one approaches the turn-off
region (see Appendix B of \citealt{Smith2009a} and the Appendix of \citealt{Smith2012}). Here also the Stripe 82 data were beneficial; the multiple epochs
provide much improved photometric accuracy compared to the rest of the SDSS
footprint (e.g. median (g-i) error was less than 10 mmag), enabling much more
precise photometric distances.

This halo sample was exploited for a variety of studies, including measuring the
tilt of the halo velocity ellipsoid out of the plane \citep{Smith2009b} and
investigating the global kinematics \citep{Smith2009a}. In the context of
this chapter, the sample was also important for investigating halo
substructures. By working in angular momentum space ($\rm L_z$, $\rm L_\perp$), the sample
shows clear evidence for the Helmi stream, for the first time detecting
candidate members up to 5 kpc away (see Fig. \ref{fig:smith}). New
candidate substructures were also found, dubbed Sloan Kinematic
Overdensities (SKOs). The first one of
these (SKOa) is particularly interesting - detected as a weak overdensity of
stars with high $\rm L_\perp$, it turned out that the feature
coincides (in angular $\rm L_z-L_\perp$ space) with a number of globular clusters taken from \citet{Dinescu1999}. Figure 10 of \citet{Smith2009a} shows how these four clusters
(NGC 5466, NGC 6934, NGC 7089/M2 and NGC 6205/M13) lie apart from the main
distribution, raising the intriguing possibility that SKOa may be either the
remains of tidal debris from one of these clusters or alternatively the remnant
of a larger galaxy which hosted a number of smaller systems (i.e. these
clusters, possibly including the progenitor of SKOa). Such hypotheses are
difficult to test, but a detailed abundance analysis will be able to
determine if its chemistry is consistent with any of these four
globular clusters.

\begin{figure}
\centering
\includegraphics[width=0.7\hsize,trim=0 0 0 0,clip]{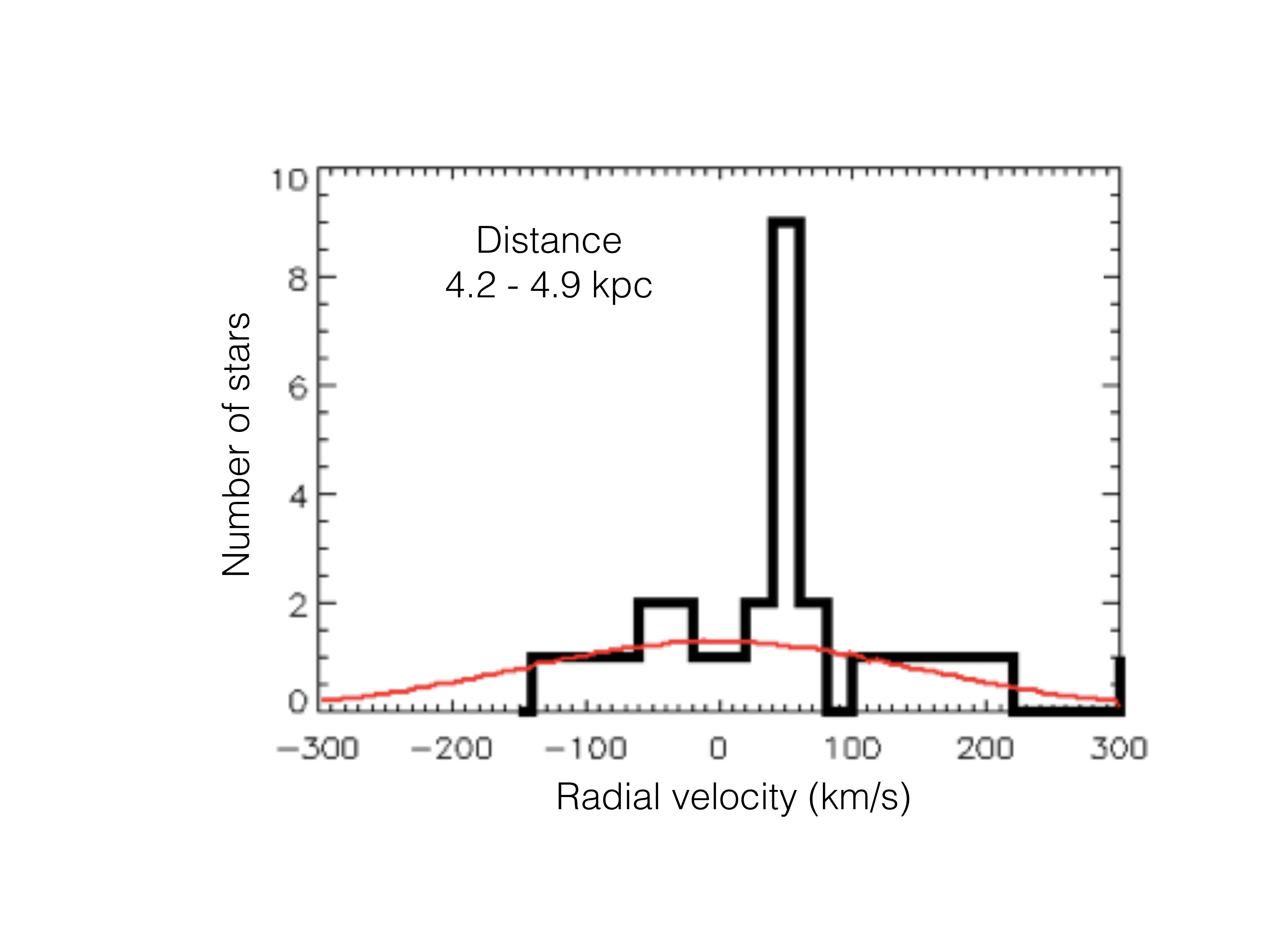}
\caption{
Confirmation of the SKOb overdensity, which was initially discovered
by \citet{Smith2009a}. This figure shows the radial velocity
distribution of stars with distances between 4.2 and 4.9 kpc,
exhibiting a clear peak at +60 km/s. The red curve denotes the
expected distribution for the smooth halo.}
\label{fig:skob}
\end{figure}

The remaining two over-densities, (SKOb \& c) arose from a search for distant
systems that are localised on the sky. The orientation of this field (it is a
long, thin stripe measuring 2.5x100 deg on the sky) means that coherent streams
are likely to ``cut through'' this narrow stripe, and so
\citet{Smith2009a} sliced the field along its length and searched for
clumping in angular momentum space. Of the two overdensities, the most
prominent is SKOb. This has been confirmed using data from MMT (Smith
et al., in prep; see Fig. \ref{fig:skob}), pinning down the distance
to between 4 and 5 kpc. This structure is intriguing because, unlike
most of the other halo moving groups mentioned in this Chapter, it
appears to be coherent, i.e. more like a stream than a moving
group. As it is close enough to measure proper motions, this means
that if we can trace an extension of this system across the sky it
will become a 6D stream, which are very rare and important for
modelling the halo \citep[e.g.][]{Koposov2010}. As with SKOa (and, for
that matter, the other overdensities found by \citeauthor{Klement2009}
in SDSS), detailed abundances would be useful to better understand the
origins of this system. However, as SDSS photometric data extends to
relatively faint magnitudes, high resolution follow-up studies are
prohibitively expensive in terms of telescope time.

\section{Distant halo streams}
\label{sec:distant}

The above discussion of local halo streams focused on those found using
three-dimensional kinematics. However, once we move beyond the solar
neighbourhood, uncertainties in proper motion prohibit the use of tangential
velocities. Tangential errors scale linearly with distance; for example, a
proper motion error of 1 mas/yr at 1 kpc corresponds to a tangential error of
4.7 km/s, but at 10 kpc this error will grow to 47 km/s. As a consequence, any
detailed structure in phase-space is washed out. Fortunately radial velocity
uncertainties do not scale with distance and so these can be used to probe
large volumes of the halo, with the caveat that it is now harder to interpret
any identified substructures, as two components of the phase-space are missing.
Also working in our favour is that in the outer halo the mixing times are much
longer, which allows ancient structures to remain coherent in configuration
space for many Gyr.

The most spectacular and important discovery of halo substructure through radial
velocities is that of the Sagittarius dwarf galaxy. For many years observers
had found dwarf galaxies around the Milky Way, beginning with the Magellanic
Clouds, but one lay hidden behind the bulge of the Milky Way and was only
discovered in 1994 \citep{Ibata1994}. This was uncovered serendipitously
during a radial velocity survey of stars towards the Galactic bulge. When
analysing the radial velocity distributions in certain fields, instead of the
expected Gaussian distribution, a secondary peak was identified. This peak
corresponded to the Sagittarius dwarf galaxy, whose systemic radial velocity is
offset from the bulge by around 150 km/s. Subsequent works have revealed that
this dwarf is in the process of being devoured by our Galaxy, with tidal
streams being discovered encircling the entire Milky Way (see Chapter 2).

At higher latitudes other streams have been identified by their
coherent radial velocities, including the Cetus
polar stream \citep{Newberg2009}, the cold metal-poor stream of
\citet{Harrigan2010} and the high-velocity stream of \citet{Frebel2013}.

\begin{figure}
\centering
\includegraphics[width=0.6\hsize]{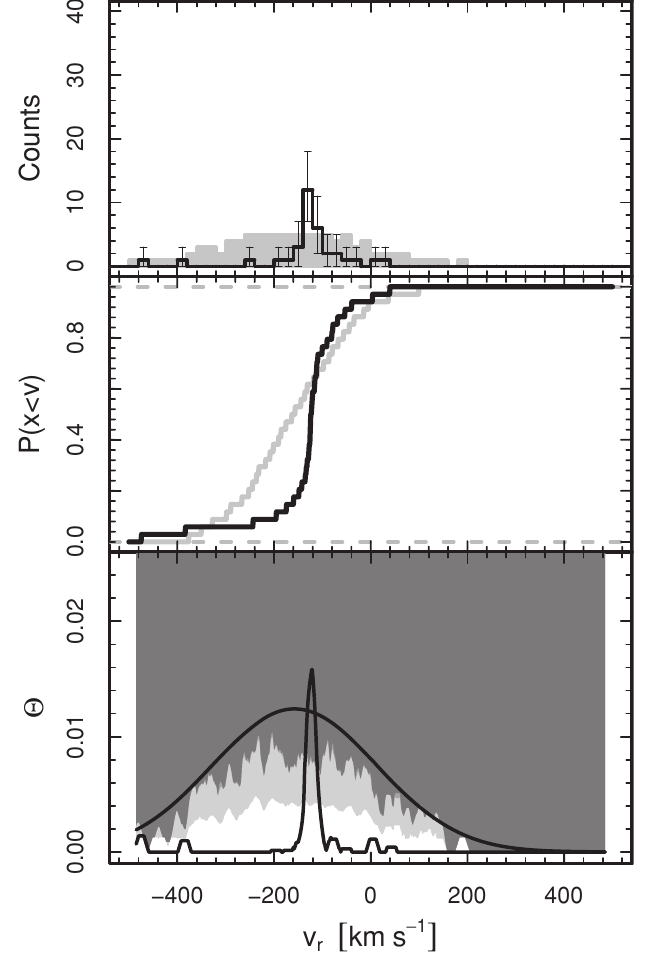}
\caption{Example of the search for an ECHOS from \citet{Schlaufman2009}. The
upper panel shows the radial velocity distribution (black histogram with error
bars) and the 95 per cent confidence interval from the multiple realisations of the
smooth halo model (grey shaded histogram). The middle panel shows the
cumulative distribution function from the observed data (black) and mean from
the realisations of the smooth halo model (grey). The lower panel shows their
measure of the statistical significance Theta (black line), with the various
significance levels denoted in light and dark grey. The wide Gaussian in this
panel shows an envelope around the most significant region, which is included
to remove any spurious detections due to small-scale fluctuations.}
\label{fig:echos}
\end{figure}

One of the most impressive studies of radial-velocity
selected substructures in the halo was led by Kevin
Schlaufman in the ``ECHOS'' series of papers
\citep{Schlaufman2009,Schlaufman2011,Schlaufman2012}.
Their approach was to take each SDSS/SEGUE spectroscopic plate and
determine, using robust statistics, whether the radial velocity distribution of
the main-sequence stars matches what one expects for a smooth halo. This was
done using two statistical tests, as described in Section 3.2 of \citet{Schlaufman2009} and illustrated in Fig. \ref{fig:echos}. The first test compared the radial
velocity histogram to a similar histogram (with the same number of radial
velocity measurements) drawn from a smooth model halo. This realization of the
smooth halo was repeatedly resampled in order to test the significance of any
peaks in the observed distribution. As can be seen from the upper panel of Fig
\ref{fig:echos}, the grey shaded region shows the 95 per cent confidence interval from these
realisations of the smooth halo model; the fact that the observation (black
histogram) and its error bar do not overlap the grey shaded region implies that
this is a robust detection. The second test is based on the cumulative
distribution of velocities, which retains more information than the previous
approach (i.e. unlike the previous approach, it avoids any binning of the
data). Again the observed distribution is compared to one drawn from a smooth
model, but this time they compare the steepness of the cumulative distribution
function. This is shown in the lower panel of Fig. \ref{fig:echos}, where one can see that
the observed slope (given by the black curve) reaches into the
high-significance regions (shown by the dark grey region).

\citeauthor{Schlaufman2009} applied these techniques to
observations of main-sequence halo stars out to 17.5 kpc on 137 individual
spectroscopic plates of SEGUE data. In these 137 lines of sight, they
identified a total of 10 strong candidates (where the number of false positive
detections is estimated to be less than 1) and a further 21 weaker candidates
(estimated to have less than 3 false positives). A number of these are likely
to be detections of existing substructures, such as the
Monoceros stream, but 7
of the strong candidates are new detections. Note that this does not translate
to 7 new independent halo substructures, as some streams could intersect
multiple lines of sight, but it does show that the halo of the Milky Way is (at
least in terms of its kinematics) lumpy, even in the inner halo where phase
mixing should occur on relatively small
time-scales. \citeauthor{Schlaufman2009} quantify this ``lumpiness,''
concluding that around 34 per cent of the inner halo is in the form of
elements of cold halo substructures (ECHOS) and estimate that there
could be as many as 1e3 individual kinematic groups in the entire inner halo.

The chemical composition of these ECHOS was
investigated in the series' second paper \citep{Schlaufman2011}.
They found that these ECHOS were more
iron-rich and less alpha-enhanced than the smooth halo, concluding that the
most-likely origin is that they were formed from the tidally disrupted debris
of relatively massive dwarf galaxies ($\rm M_{tot} > 1e9 ~M_\odot$). In the
final paper of this series \citep{Schlaufman2012} the authors
investigated the spatial coherence
in [Fe/H] as a function of Galactocentric radius, concentrating only on main
sequence turnoff stars from the kinematically smooth halo (namely SEGUE fields
in which no ECHOS were detected). Although these fields are phase mixed and
show no kinematic substructures, the chemistry of these stars can illuminate
their origins. By studying the distributions of stars in the $\rm
[Fe/H]$--$\rm [\alpha/Fe]$ plane they found that the accreted halo becomes dominant beyond around 15
kpc from the Galactic centre, arguing that at smaller radii the halo is
probably formed from a combination of in-situ star formation and dissipative
major mergers at high redshift.

As the careful work of Schlaufman et al. has shown, statistical studies are
important if we are to dissect the halo of the Milky Way beyond the solar
neighbourhood, especially when the sky coverage is not contiguous and the
sampling of stars with spectra is sparse. Various tools have been developed,
including the 4distance measure introduced by Starkenburg et al. (\citeyear{Starkenburg2009}; see
also the similar approach of \citealt{Clewley2006}). This technique, which is
based around the separation of pairs of stars in four dimensions (angular
position on the sky, distance and radial velocity), has proved influential for
subsequent works \citep[e.g.][]{Cooper2011,Xue2011} and will
undoubtedly continue to be used on surveys of distant halo stars where proper
motions are unavailable.

\section{Future prospects}
\label{sec:future}

Despite extensive progress in identifying kinematic streams, the field is far
from exhausted. On the contrary, this decade is likely to see a resurgence in
this field, leading to unprecedented insights into the formation of our
Galaxy.

From the current generation of spectroscopic surveys, we can expect significant
progress in the coming years. In terms of sheer volume of spectra, the Chinese
LAMOST survey is unsurpassed \citep{Deng2012}. By gathering over a million
spectra each year, this spectroscopic survey has great potential. Already one
new candidate kinematic overdensity has been identified \citep{Zhao2014} and
various works are analysing substructure in the local velocity distribution,
for example the work of \citet{Xia2015} which is utilising the extreme
deconvolution technique \citep{Bovy2011}.

If we think about the accreted galaxies which built up our stellar halo, they
will of course have a range of masses and accretion times. Their chemical
composition will therefore vary since the amount of enrichment that can take
place depends on these factors (see, for example, \citealt{Lee2015}). As a
consequence, a detailed dissection of the accretion history of our halo will
require both kinematics and chemistry. By combining dark matter simulations
with semi-analytic prescriptions for the star formation and chemistry, it is
possible to make predictions for what we may be able to detect and how much we
can infer about our Galaxy's accretion history from a given set of of kinematic
and chemical abundance data \citep[e.g.][]{Johnston2008}.

There are a number of spectroscopic surveys that operate at resolutions
sufficient to carry out detailed chemical abundance analyses, for example the
SDSS project APOGEE \citep{Holtzman2015}, the GALAH survey
\citep{Freeman2012}, or the Gaia-ESO survey
\citep{Gilmore2012,Randich2013}. These detailed
abundances opes up the possibility of ``chemical tagging,'' whereby abundance
ratios are used to disentangle the different formation sites for groups of
stars \citep{Freeman2002}. Clearly these additional dimensions
will be extremely valuable when attempting to identify kinematic substructures
in the local disk, where groups may overlap if one looks at only the 6D
phase-space. Although this technique is ideally suited to finding moving groups
in the disk, as mentioned above chemistry will allow us to classify halo
streams and understand their origins -- in effect carrying out the
discovery and follow-up in one step.

As the Gaia satellite begins to deliver scientific
return, there is no doubt that we are on the cusp of a true revolution in this
field. This mission, which is led by the European Space Agency, is collecting
high precision astrometry of a billion stars in our galaxy. All stars in the
sky brighter than 20th magnitude will be observed, leading to exquisite proper
motions and parallaxes. The precision is so great that it will be able to
measure distances (through trigonometric parallax) to less than 1 per cent for
ten million stars. In addition to the astrometry, Gaia will provide detailed
photometric information (from spectrophotometry)
including stellar parameters
and, for stars brighter than around 17th magnitude,
spectroscopic information including radial velocities. A description of the
science capabilities can be found in \citet{deBruijne2012}, although
continually updated performance information can be found on the Gaia 
\href{http://www.cosmos.esa.int/web/gaia/science-performance}{webpage}.
The final catalogue is expected in 2022, with
interim releases before then.

\begin{figure}
\centering
\includegraphics[width=0.8\hsize]{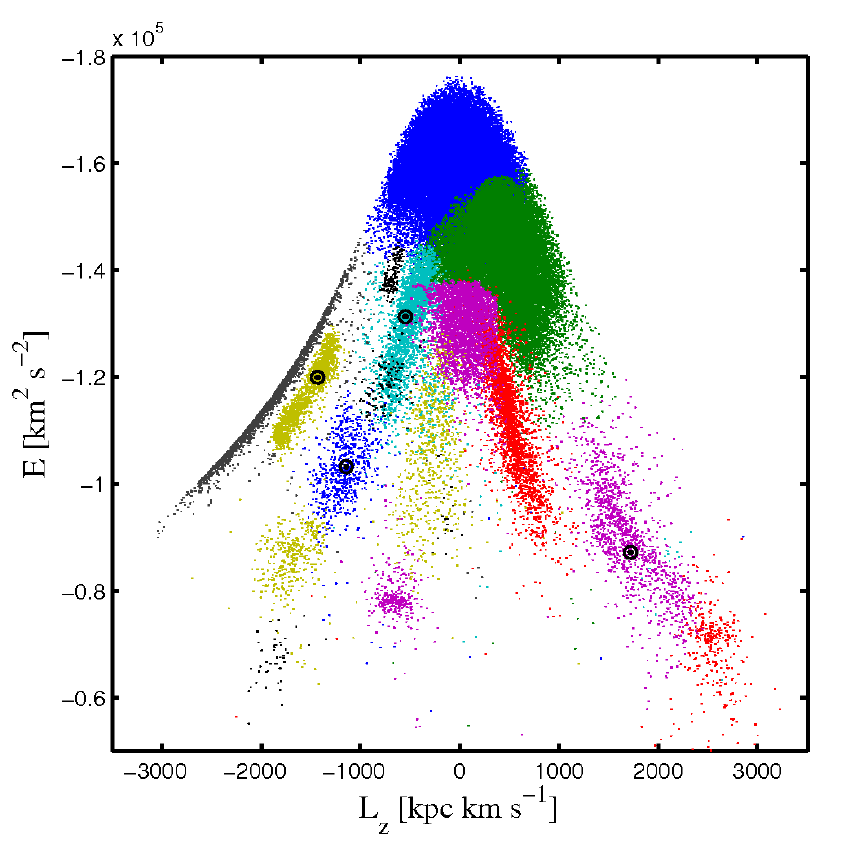}
\caption{
An example of how Gaia might see the distribution of accreted
satellites in the solar neighbourhood. The different colours correspond to
different satellites in the space of energy and the vertical component of the
angular momentum. Black circles denote the four satellites for which
it will be possible to estimate the time of accretion. Taken from
\citet{Gomez2010a}.}
\label{fig:gaia}
\end{figure}

Clearly such an unprecedented mapping of 6D phase space will open up an entirely
new view of the local velocity distribution. While we wait for the first Gaia
data to appear, many authors have attempted to estimate what we might be able
to see. One example of this is \citet{Gomez2010a}, who modelled the Milky Way
halo through the accretion of satellite galaxies, then convolved these with
Gaia's observational errors. Fig. \ref{fig:gaia} shows what we may be able to detect in a
solar neighbourhood realization; there are 1e5 stellar halo particles within
this sphere of 4 kpc radius, plus around 20,000 stellar disk particles. Upon
applying a detection algorithm to identify substructure, they confirm 12
separate accretion events, corresponding to around 50 per cent of all disrupted
satellites in this volume. For some of these detections the authors find that
it should be possible to directly estimate when these satellites were accreted,
exploiting the fact that disrupting satellites form separate clumps in
frequency space and the separation of these clumps relate to the time since
accretion \citep{McMillan2008,Gomez2010b}. This remarkable feat
requires a large enough sample of stars with accurate parallaxes (typically 50
or more stars with parallax error less than 2 per cent), but in this
realization \citeauthor{Gomez2010a} predict that it should be attainable for at least four
of their detected satellites. Being able to determine the time of accretion,
together with a detailed analysis of the chemistry of these stars, will
undoubtedly teach us a great deal about the evolution of star formation in
these earliest galaxies.

With Gaia in mind, a number of other studies have devised methods to search for
substructures. One such work is that of \citet{Mateu2011}, who utilise the
fact that (for a spherical potential) streams will fall on great circles as
viewed from the Galactic centre. This is based on an earlier study \citep{Johnston1996}, but by extending the analysis to include data such that will be
available from Gaia (i.e. parallaxes and kinematics) the method has much
greater efficacy. Of course this technique still requires streams to be
confined to orbital planes and, as such, is ill-suited to the inner halo where
the shorter dynamical times lead to significant phase mixing. However, at
intermediate distances in the halo where Gaia will still be able to provide
reasonable parallaxes (with distance accuracy of say 30 per cent), this technique will
thrive.

Although Gaia will play a dominant role in the coming decades, it will not
provide all of the answers. The on-board spectroscopy is limited to only the
brightest stars and will not deliver detailed chemistry, meaning that a huge
ground-based follow-up program is required. The realization that this
limitation hampered the scientific return of the Hipparcos mission, led to the
Gaia-ESO survey, and also provides strong motivation for future instruments,
such as Subaru's Prime Focus Spectrograph \citep{Takada2014}, \ 4-MOST \citep{deJong2014}, WEAVE \citep{Dalton2014} and the Maunakea Spectroscopic
Explorer \citep{Simons2014}.

It is fascinating to see how, 150 years since \citeauthor{Madler1846}
and his contemporaries made their first discoveries, the analysis of
kinematic substructures is still playing an important role in
understanding the evolution of the Galaxy. M{\"a}dler couldn't have
imagined that some moving groups could be the relics of other 
galaxies, but today these substructures are illuminating our knowledge of the
earliest galaxies and hierarchical assembly. As new surveys are undertaken, the
census of substructures becomes more complete. Perhaps in 150 years accretion
events such as these will still be contributing new insights.

\begin{acknowledgement}
The author acknowledges financial support from the CAS One Hundred Talent Fund,
NSFC grants 11173002 and 11333003, the National Key Basic Research Program of
China (2014CB845700) and the Strategic Priority Research Program ``The
Emergence of Cosmological Structures'' of the Chinese Academy of Sciences
(XDB09000000). This work is partially supported by the Gaia Research for
European Astronomy Training (GREAT-ITN) Marie Curie network, funded through the
European Union Seventh Framework Programme (FP7/2007-2013) under grant
agreement No. 264895. This chapter uses data obtained through the Telescope
Access Program (TAP), which has been funded by the Strategic Priority Research
Program ''The Emergence of Cosmological Structures'' (Grant No. XDB09000000),
National Astronomical Observatories, Chinese Academy of Sciences, and the
Special Fund for Astronomy from the Ministry of Finance.
\end{acknowledgement}

\clearpage

\bibliography{smith}

\end{document}